\documentclass[iop]{emulateapj}

\usepackage{natbib} 
\usepackage[caption=false]{subfig} 

\newcommand{\emailpazch}{francisco@dfte.ufrn.br} 
\newcommand{\teff}{$T_{\rm eff}\,\,$}
\newcommand{\logg}{$\log g\,\,$} 
\newcommand{\period}{$P_{\rm rot}\,\,$}
\newcommand{\msun}{M$_{\sun}\,\,$}

\slugcomment{Submitted to Astrophysical Journal, \today} 
\shorttitle{Rotational behavior \emph{Kepler} Host Stars}
\shortauthors{Paz-Chinch\'{o}n et al.}

\begin{document}

  \title{ {The rotational behavior of \emph{Kepler} Stars with Planets} }

  \author{F. Paz-Chinch\'{o}n$^ {1,4,\dagger}$, I. C. Le\~{a}o$^ {1}$, J.~P. Bravo$^ {1}$, D.~B. de Freitas$^ {1}$, C.~E. Ferreira Lopes$^ {1,3}$, S. Alves$^ {2,}$   \altaffilmark{5}, M. Catelan$^ {2,4,}$\altaffilmark{6}, B.~L. Canto Martins$^ {1}$, and J.~R. De Medeiros$^ {1}$}

  \affil{$^{1}$Departamento de F\'{i}sica Te\'{o}rica e Experimental, Universidade Federal do Rio Grande do Norte, Campus Universit\'{a}rio, Natal, RN, Brazil}
  \affil{$^ {2}$Instituto de Astrof\'{i}sica, Pontificia Universidad Cat\'{o}lica de Chile, Av. Vicu\~{n}a Mackenna 4860, Macul, Santiago, Chile}
  \affil{$^ {3}$SUPA Wide--Field Astronomy Unit, Institute for Astronomy, School of Physics and Astronomy, University of Edinburgh, Royal Observatory, Blackford Hill, Edinburgh EH9 3HJ, UK}
  \affil{$^ {4}$Millennium Institute of Astrophysics (MAS), Santiago, Chile}
  \altaffiltext{5}{CAPES Foundation, Ministry of Education of Brazil, Bras\'{i}lia, DF, Brazil}
  \altaffiltext{6}{Centro de Astro--Ingenier\'{i}a, Pontiﬁcia Universidad Cat\'{o}lica de Chile, Santiago, Chile}
  \altaffiltext{$\dagger$}{Correspondence author: \emailpazch}

\begin{abstract}

We analyzed the host stars of the present sample of confirmed planets detected by \emph{Kepler} and 
\emph{Kepler} Objects of Interest (KOI) to compute new photometric rotation periods and to study the 
behavior of their angular momentum.  Lomb--Scargle periodograms and wavelet maps were computed for $3\!,807$ stars. 
For 540 of these stars, we were able to detect rotational modulation of the light curves at a significance 
level of greater than 99\%. For 63 of these 540 stars, no rotation measurements were previously available 
in the literature. According to the published masses and evolutionary tracks of the stars in this sample, 
the sample is composed of M-- to F--type stars (with masses of 0.48--1.53~M$_{\sun}$) with rotation periods that span 
a range of 2 to 89~days. These periods exhibit an excellent agreement with previously reported (for the stars 
for which such values are available),  and the observed rotational period distribution strongly agrees with theoretical 
predictions. Furthermore, for the 540 sources considered here, the stellar angular momentum provides an important 
test of Kraft's relation based on the photometric rotation periods. Finally, this study directly contributes in a direct 
approach to our understanding of how angular momentum is distributed between the host star and its (detected) planetary 
system; the role of angular momentum exchange in such systems is an unavoidable piece of the stellar rotation puzzle.

\end{abstract}


\section{Introduction}

The study of stellar rotation is important for our understanding of many 
kinematic properties of stars, including the origin and evolution of the rotation itself;
more generally, the origin and evolution of stellar angular momentum; and the role of 
rotation in stellar magnetism and mixing of chemical elements. Stars are formed from
rotating molecular cloud cores and preserve only a very small fraction of the initial
angular momentum of these cores \citep[e.g.,][]{palla-02,lamm-05}. Indeed, the initial angular momentum
of the initial molecular cloud is approximately eight orders of magnitude greater than that of the
stars that eventually form from such a cloud core \citep[e.g.,][]{bodenheimer-89}. With
the discovery of extra--solar planets \citep[e.g.,][]{mayor-95}, new questions concerning
rotation have begun to arise, including questions regarding the large--scale effects of the presence
of planetary companions on stellar rotation \citep{pont-09,alves-10,lanza-10} and
on the heating of stellar coronae and chromospheres
\citep[e.g.,][]{poppenhaeger-10,cantomartins-11,lanza-12}. Admittedly, the study of the
rotation of planet--hosting stars is hampered by one major factor: for the great majority
of stars with planets only the projected rotational velocity $v\ \sin i$ is available, and all such 
bodies identified, to date are essentially slow--to--very--moderate rotators \citep{alves-10}. Despite 
this constraint, Alves et al. have noted that stars without any detected planets demonstrate a
clear angular momentum deficit compared with stars with planets.

The recent space--missions CoRoT \citep{baglin-07} and \emph{Kepler}
\citep{koch-10} are providing a new window into the study of the behavior
of stellar rotation, including the rotation of stars with planets, by virtue of their high--precision
photometry, which yields excellent quality light curves (LCs). In addition to their high
precision, these missions provide nearly uninterrupted photometric measurements of
unprecedented duration and cadence, allowing for refined analysis of the features involved in the 
behavior of these temporal series, as in \citet{defreitas-13b}.
In recent works, \citet{mcquillan-13a,mcquillan-13b,mcquillan-14}, \citet{nielsen-13}, \citet{walkowicz-13}, 
and \citet{reinhold-13} computed the rotation periods ($P_{\rm rot}$) for a large fraction of the 
\emph{Kepler} targets, including the \emph{Kepler} Objects of Interest (hereafter KOI).  
\citet{mcquillan-13a,mcquillan-13b,mcquillan-14} analyzed the first 11 quarters of \emph{Kepler} data by
computing Fourier periodograms and autocorrelation functions (ACF), and validated their results through visual inspection. 
\citet{mcquillan-13a} computed the rotation periods of $1\!,570$ M dwarfs and identified a possible change 
in the slope of the period--mass relation at approximately 0.55~M$_{\sun}$. 
\citet{mcquillan-13b} measured the periods of 737 main--sequence KOI stars to study the relation between 
the rotational and orbital periods; they found that hotter stars rotate more rapidly.
\citet{mcquillan-14} derived periods for $34\!,030$ main--sequence stars, excluding
eclipsing binaries and KOI stars, and found that the majority of these stars are typically
younger than 4.5~Gyr and exhibit higher peak--to--peak amplitudes for smaller periods.
Moreover, \citet{nielsen-13}, \citet{reinhold-13}, and \citet{walkowicz-13} computed periods
automatically using Lomb--Scargle periodograms. \citet{nielsen-13} reported rotation
periods for $12\!,151$ stars.  \citet{reinhold-13} derived periods for $24\!,124$
targets using an automatic method with partial visual inspection. These authors identified a trend of 
magnetic braking when comparing {rotation} periods with the gyrochronological relations from \citet{barnes-07}.  
\citet{walkowicz-13} reported rotation periods and ages for approximately 950 KOI stars, and through comparison with
spectroscopic data, they estimated the corresponding stellar orientation angles. These authors also used 
gyrochronological ages based on \citet{barnes-07}, in combination with the calibrations of
\citet{mamajek-08} to estimate stellar ages.

In this work we present rotation periods for 131 planet--hosting \emph{Kepler}
stars. We computed the rotation periods using a robust procedure, that combines the standard
Lomb--Scargle periodogram \citep{lomb-76,scargle-82} with the wavelet technique
\citep{grossmann-84}.  We also present rotation periods for a comparison sample of
 409 KOI stars (released December 13, 2013). By combining these data with
theoretical predictions, we are able to present an overview of the rotational behavior of
these stars {on} the Hertzsprung--Russell (HR) diagram.  This work is organized as follows:
in Section \ref{KeplerData}, the working stellar sample, the \emph{Kepler} data and the
procedure used for the computation of periods are presented.  Section \ref{Results}
presents the results. Our conclusions are presented in Section \ref{Conclus}.

\section{Kepler Data and Analysis} \label{KeplerData}

From May 2009 to May 2013 the NASA \emph{Kepler} mission collected data in an
steady field of view for $191,\!449$ stars in 17 runs (known as
\textit{quarters}) which were composed of \textit{long cadence} \citep[6.02 s observations 
stacked every 29.4 minutes,][]{jenkins-10b} and \textit{short cadence} (bins of 59
seconds) observations \citep{DR-Q1,DR-Q17}.  
We selected all public light curves LCs available in the \emph{Kepler} Public 
Archive\footnote{\anchor\url{http://archive.stsci.edu/kepler/}} for
planet--hosting stars and KOIs between quarters 1 and 17, which were collected over the noted
range of four years (from May 13, 2009, to May 08, 2013).  From this parent sample, 408
\emph{Kepler} Confirmed Planetary Host
Stars\footnote{\anchor\url{http://kepler.nasa.gov/Mission/discoveries/}} (hereafter
KCP) were available in the \emph{Kepler} Public Archive.  Details regarding the public data
have been discussed in many publications
\citep[e.g.,][]{borucki-09,borucki-10,batalha-10,koch-10,basri-11}.  In particular, the
\emph{Kepler} database provides both Simple Aperture Photometry (SAP) data, which were processed
using a standard treatment, and Pre--Search Data Conditioning (PDC) data, which were analyzed using
a more refined treatment based on the PDC routine from the \emph{Kepler} pipeline
\citep{jenkins-10a}. The PDC routine primarily removes thermal and kinematics effects caused
by the spacecraft \citep[see][for further details]{handbook}.

\begin{figure}[t!]
  \plotone{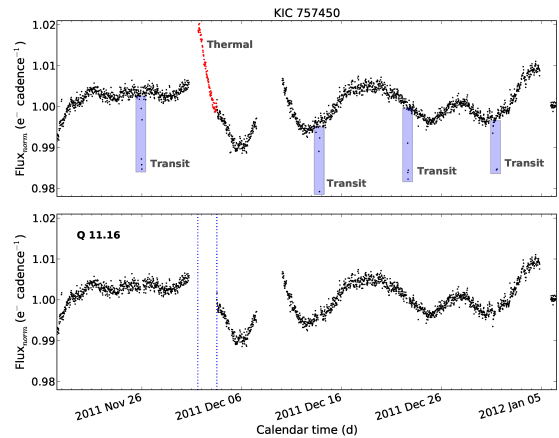}
  \caption{Example of a LC before (upper panel) and after (bottom panel) the
      removal of transits and thermal noise.}
  \label{thermal}
\end{figure}

In this work, the parent samples of the KCP host stars and KOIs were analyzed using the PDC LCs. 
The KOIs were used as a comparison sample, because of the large number of stars with this classification 
(almost all KCP host stars are also KOIs). Most of these LCs are of excellent quality and required no 
additional treatment of artifacts.  However, for the final solutions some \textit{flare--like}\footnote{We named
\textit{flare--like} signature a sudden and strong flux bump in the LC in short time
interval (typically of few days), whose physical or instrumental origin is not
important to this work.} signatures  \citep[as described by][]{maehara-12} and
known planetary transits were removed.
Removing transits and \textit{flare--like} signatures allows for better isolation of other
sources of stellar LC variability, which may include rotational modulation. In total we
removed all transits from all 408 stars of the KCP parent sample and 61
\textit{flare--like} signatures found in 26 (of the 408) stars of the KCP parent sample.
We also removed outliers by discarding all flux measurements that exceeded 3.5\,$\sigma$
from a 3$^{rd}$--order polynomial fit to each quarter\textquoteright s LC.  Furthermore, 
some few remaining instrumental trends \citep[as described by][]{petigura-12} were
removed. These included, in particular, thermal noise (which was predominantly produced by
variations in the overall spacecraft temperature that were induced by the monthly pointing of the antenna
toward Earth for data transmission), which manifested as exponential decays in the LCs
\citep[see][]{petigura-12,garcia-11}, and \textit{safe--mode} signatures (induced by
cosmic rays hitting the detector), which also exhibited high--slope decays \citep{garcia-11}.
Both types of artifact exhibited very similar shapes, and they were visually identified for exclusion.
Figure~\ref{thermal} presents an example of a LC from which transits and artifacts were
removed. Because both the Lomb--Scargle periodograms and wavelet maps were calculated using 
discrete transforms of the time series, both types of analysis were capable of treating the 
gaps introduced by removing the data, as described above.  
Finally, individual quarters were combined to produce a single long--term LC for each
object.  The quarter--to--quarter transitions typically manifested as flux offsets (jumps),
which were adjusted through a linear fit and extrapolation of user--defined boxes before and
after each jump, as described by \citet{demedeiros-13} and \citet{banyai-13}. 

\begin{figure*}[h!bp]
  \centering
  \subfloat{ \label{pwv:1}\includegraphics[scale=0.32]{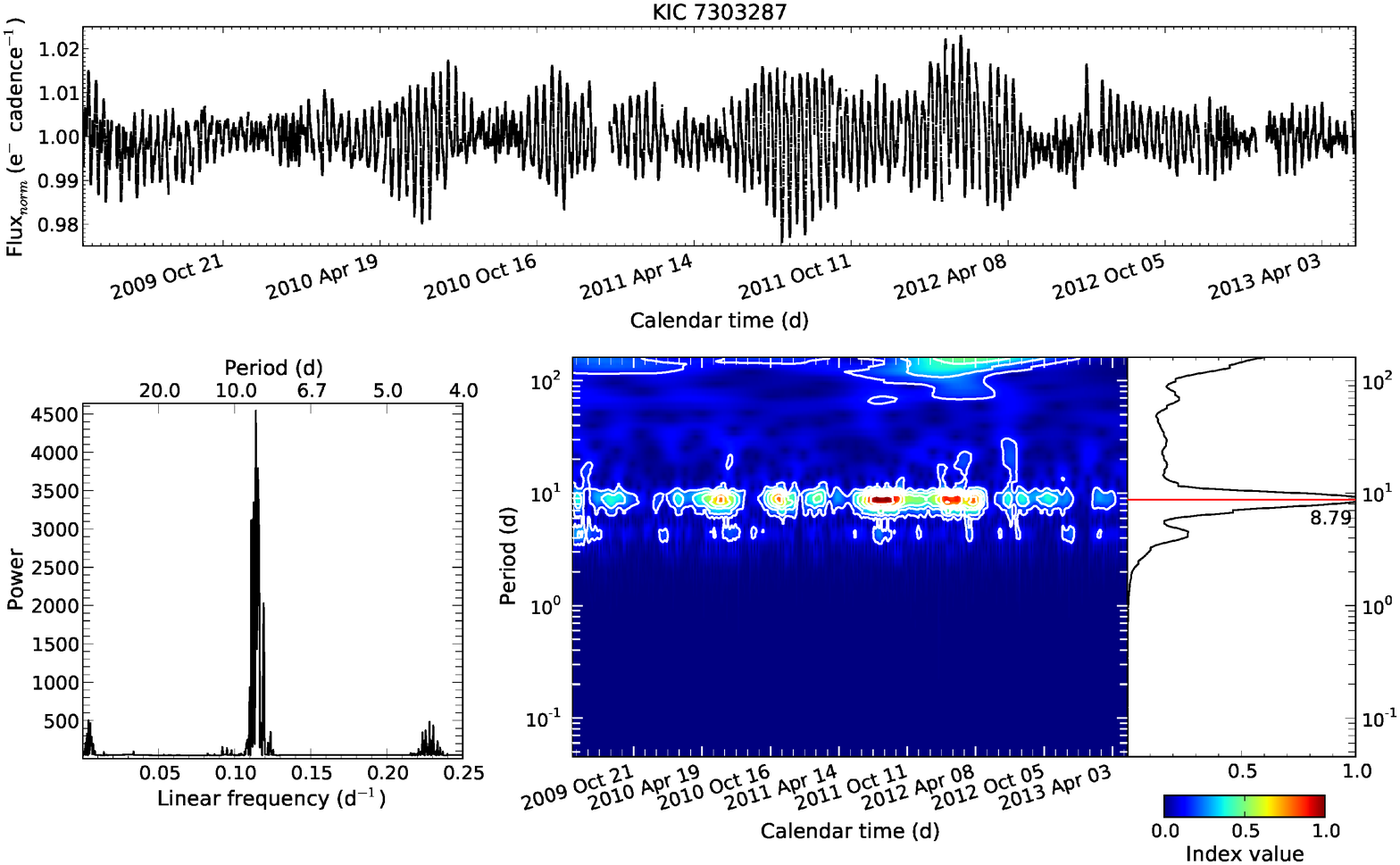} } 
  \,
  \subfloat{ \label{pwv:2}\includegraphics[scale=0.32]{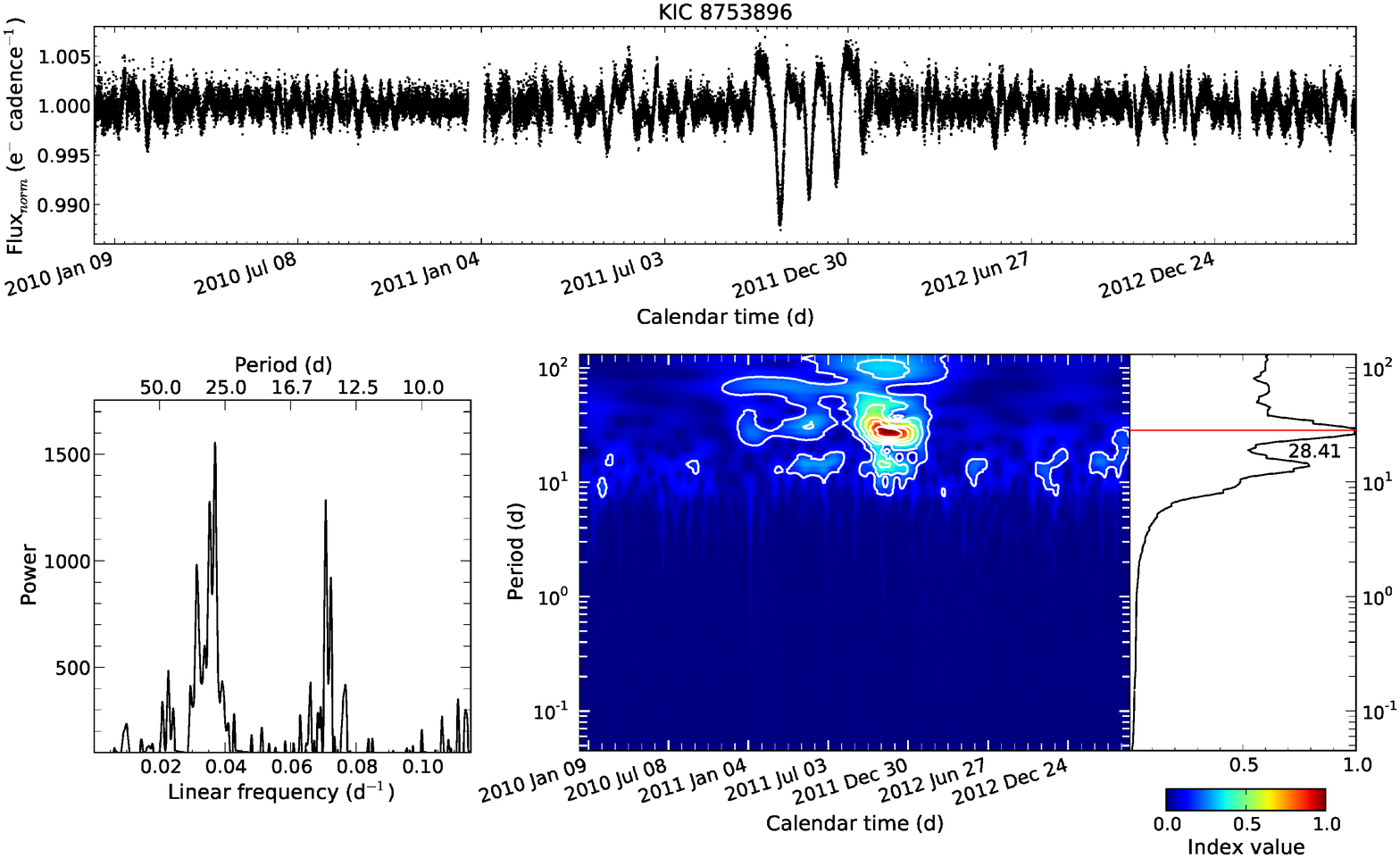}}
  \, 
  \subfloat{ \label{pwv:3}\includegraphics[scale=0.32]{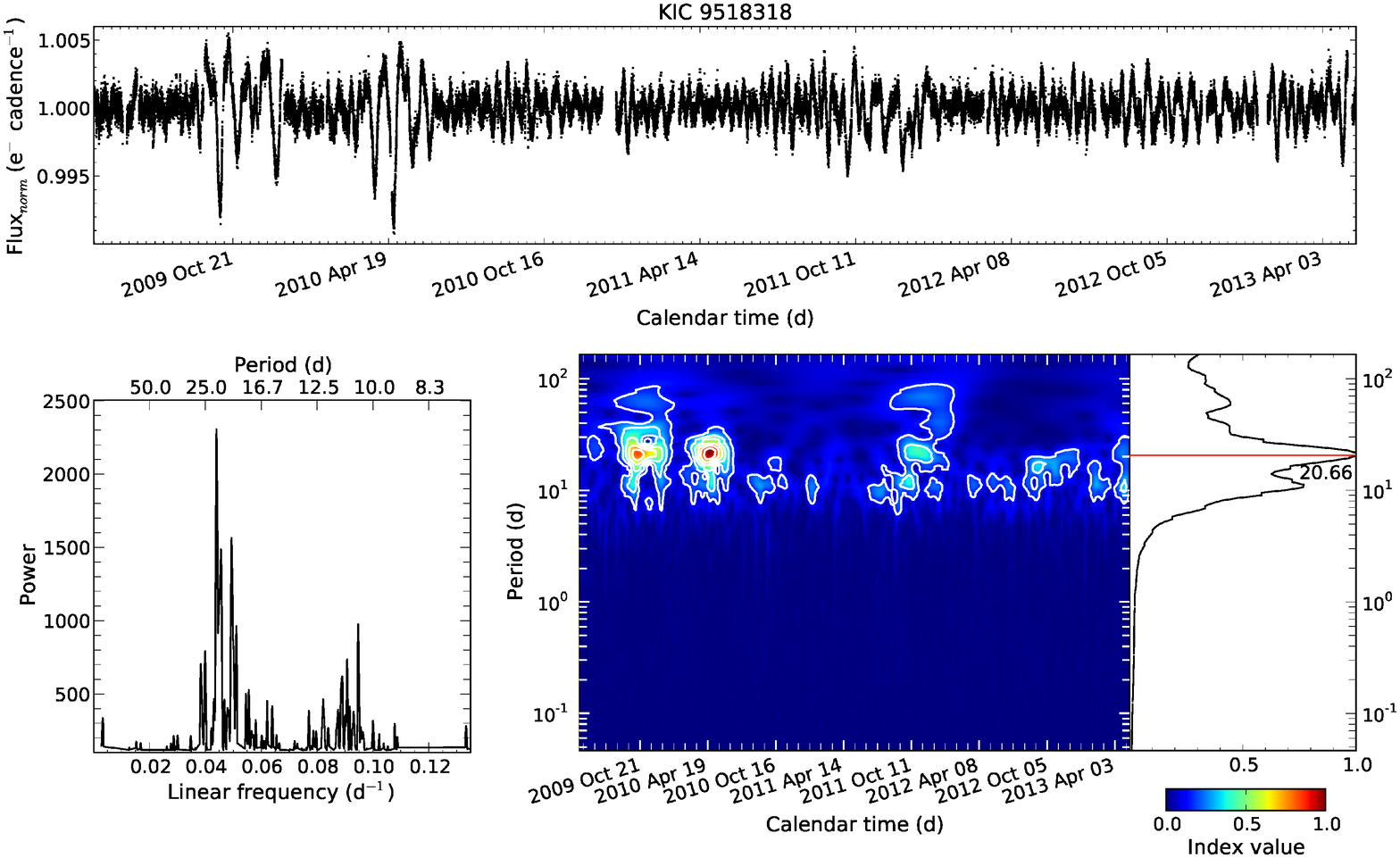}}
  \,
  \caption{{Continued on} the next page.}
\end{figure*}
	
\begin{figure*}[h!tp]
  \ContinuedFloat 
  \centering 
  \subfloat{ \label{pwv:4}\includegraphics[scale=0.32]{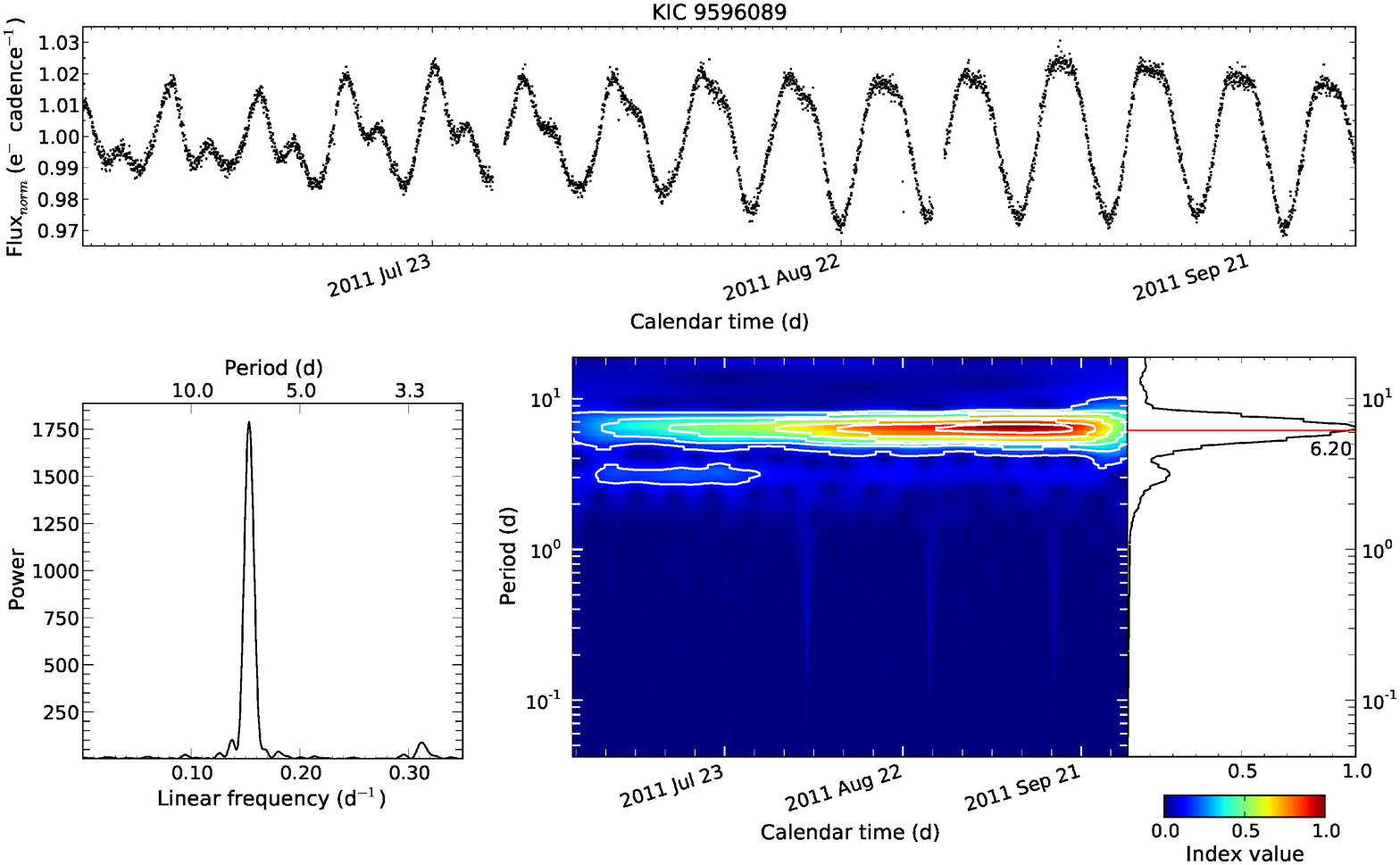} } 
  \,
  \subfloat{ \label{pwv:5}\includegraphics[scale=0.32]{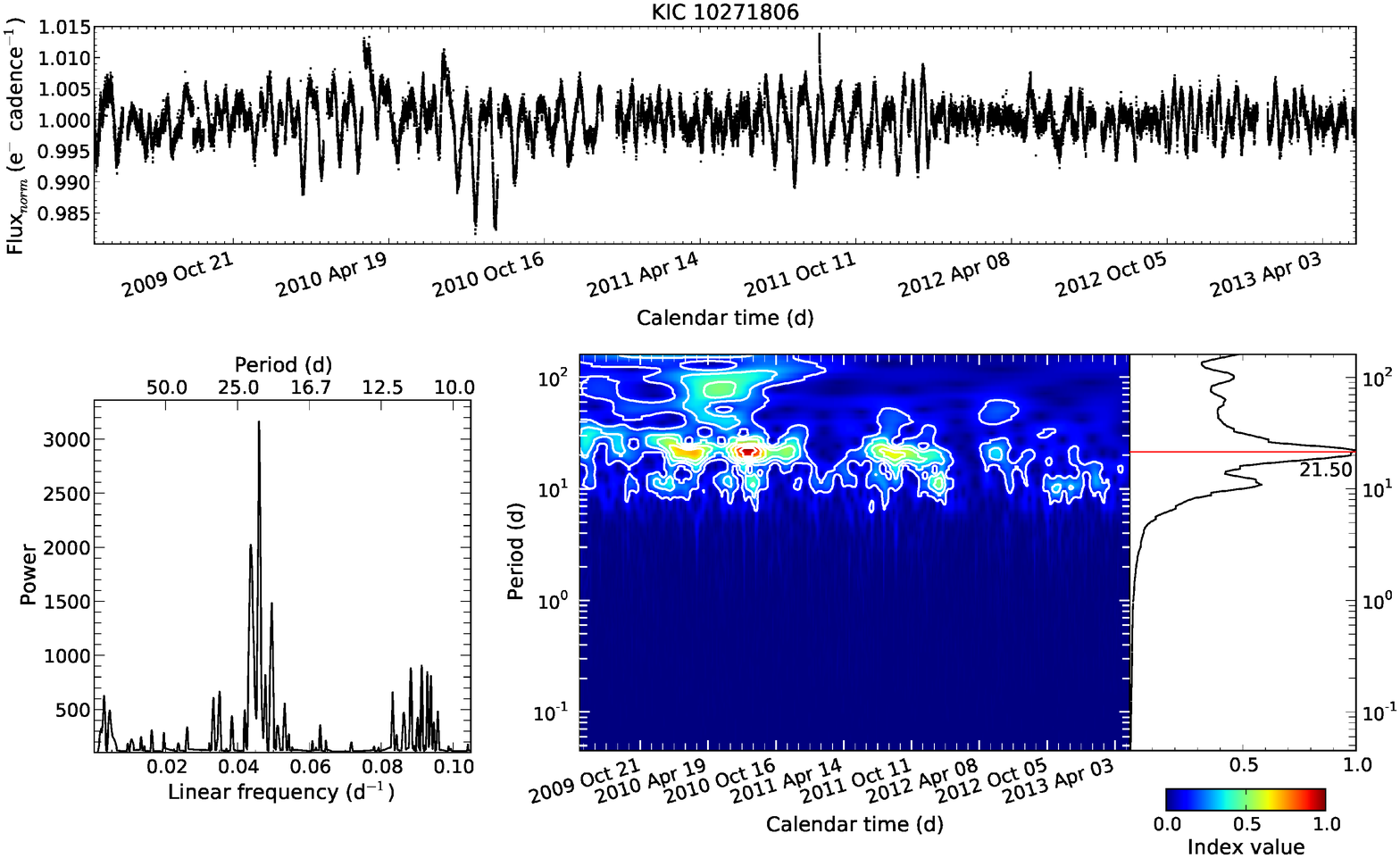}}
  \, 
  \subfloat{ \label{pwv:6}\includegraphics[scale=0.32]{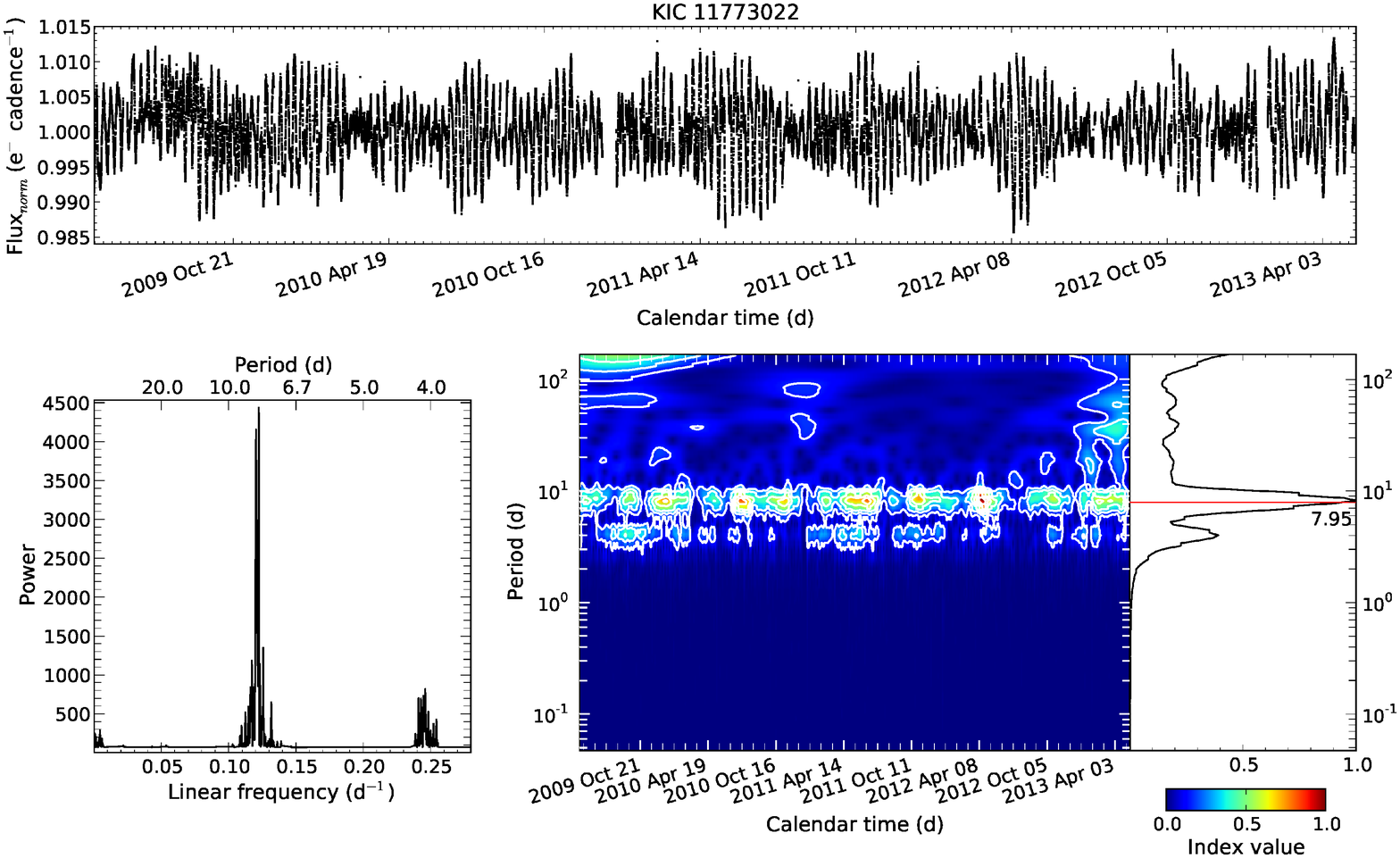}}
  \,
  \caption{Typical periodograms and wavelet maps for several stars from our final
      sample. Each panel is configured as follows: at the top is the LC (distinguished 
      by its KIC identifiers),  
      at the bottom left is the Lomb--Scargle periodogram; 
      and at the bottom right are the wavelet map and the global spectrum.
      In the Lomb--Scargle periodograms, the power corresponding
      to a false alarm probability (FAP) of 0.01 is not displayed because 
      it is visually very close to the $x$--axis.      
      }
    \label{wvm:gral}
\end{figure*}

For data analysis, Lomb--Scargle periodograms were computed over the entire
sample within a frequency range of 0.01 -- 0.5~d$^{-1}$.  As in \citet{reinhold-13}, we
defined a maximum frequency of 0.5~d$^{-1}$ (or a minimum period of 2.0~d) to avoid pollution 
of the samples with sources of variability other than rotational modulation (typically pulsations or oscillations).  
The Lomb--Scargle method was based on the approach of \citet{press-89} and standard statistical methods
\citep[e.g.,][]{scargle-82,horne-86,press-89} were used to compute confidence levels.
Accordingly, in each periodogram, we identified the main peak with a confidence level of
greater than 99\%.  We also computed wavelet maps and global wavelet power spectra
within the same frequency range that was used for the Lomb--Scargle periodograms, following the
procedure described in \citet{bravo-14}.  These authors treated the 6$^{th}$--order
Morlet wavelet as the mother wavelet, because of its good time localization and frequency
resolution \citep{grossmann-84}. The width of the mother wavelet was variable and was
defined by $s_i = s_0 2^{n_{i} \delta}$ \citep{torrence-98}, where $s_i$ is the
so--called \textit{scale} (the period decomposition in our case), $s_0$ is the minimum
scale ({with a} value of 2.0 for our calculations), $\delta$ is the scale bin (with a value of 0.1 for
our calculations), and $n_{i}$ is an iterative index that ranges from 0 to $N-1$, where $N$ is
the total number of scales. Wavelet analysis is a powerful tool for analysis of
frequency variations over time, for a given signal decomposed at various
resolutions. In addition to revealing the stability of superposed
variability periods over time, the time--frequency representation of wavelet maps allows
us to identify specific signatures that are related to variability behaviors.  Typical
examples of the wavelet maps computed in our analysis are displayed in Fig.~\ref{wvm:gral},
together with their corresponding LCs. For the statistical analysis of global wavelet
periodograms, we followed the approach presented in \citet{torrence-98} (see \S~{4}
and Eq.~20 {in the cited publication}).  We then identified the main peaks in the global wavelet 
periodograms with confidence levels of greater than 99\%.

Subsequently, we performed an automatic pre--selection of KOIs and KCP host stars, which were 
then visually inspected for our final selection. This automatic pre--selection was less strict for 
KCP host stars than for KOIs because the KCP host stars comprised our main sample, which was sufficiently 
small for visual inspection to be feasible.  Below we describe the procedures used in our pre--selection 
for the KCP and KOI subsamples.  Finally, we estimated the stellar angular momentum of each target 
(the results are presented in \S~\ref{Results}) and also estimated the uncertainties of all computed parameters.
These final steps are described below.  The computed periods, with values that ranged from 2.64 -- 36.25~d 
(KCP host stars), and 2.05 -- 88.94~d (KOIs), are available in Table~\ref{Table}.

\begin{deluxetable*}{rccccccccc}
  \tabletypesize{\scriptsize} 
  \tablewidth{0pt}
  \tablecaption{A sub--sample of the final KCP sample with the rotation periods computed in
    this work\tablenotemark{a}.}
  \tablecolumns{9} 
  \tablehead{ 
    \colhead{KIC} & \colhead{$P_{\rm rot}$\tablenotemark{b}} &
    \colhead{Amplitude\tablenotemark{b}} & \colhead{$T_{\rm eff}$\tablenotemark{c}} &
    \colhead{$\log g$\tablenotemark{d}} & \colhead{M$_{star}$\tablenotemark{e}} &
    \colhead{R$_{star}$\tablenotemark{e}} & \colhead{S/N} & 
    \colhead{Flag\tablenotemark{f}} &
    \\ 
    \colhead{} & \colhead{(d)} & 
    \colhead{(e$^{-}$ cadence$^{-1}$)} & \colhead{(K)} &
    \colhead{(cm s$^{-2}$)} & \colhead{(M$_{\sun}$)} &
    \colhead{(R$_{\sun}$)} & \colhead{} &
    \colhead{} 
  } 
  \tablecolumns{9} 
  \startdata 
   757450 & 19.19\ $\pm$\ 2.39 & 0.00383\ $\pm$\ 0.00296 & 5152\ $\pm$\ 93 & 4.54\ $\pm$\ 0.30 & 0.88 & 0.88 & 8.0 & 1 \\
  2165002 & 22.59\ $\pm$\ 2.03 & 0.00147\ $\pm$\ 0.00214 & 5245\ $\pm$\ 83 & 4.57\ $\pm$\ 0.30 & 0.8 & 0.79 & 2.89 & 1 \\
  2302548 & 12.35\ $\pm$\ 2.14 & 0.00694\ $\pm$\ 0.00297 & (...) & (...) & 0.87 & 0.79 & 34.81 & 1 \\
  2438264 & 17.93\ $\pm$\ 1.87 & 0.00032\ $\pm$\ 0.00108 & 5031\ $\pm$\ 74 & 4.57\ $\pm$\ 0.30 & 0.74 & 0.81 & 1.52 & 1 \\
  2692377 & 21.51\ $\pm$\ 1.24 & 0.00049\ $\pm$\ 0.00115 & (...) & (...) & 0.99 & 1.11 & 4.96 & 2 \\
  3323887 & 16.81\ $\pm$\ 0.69 & 0.00064\ $\pm$\ 0.00132 & (...) & (...) & 1.07 & 1.02 & 4.08 & 1 \\
  3541946 & 15.21\ $\pm$\ 0.86 & 0.00016\ $\pm$\ 0.00370 & 5725\ $\pm$\ 71 & 4.56\ $\pm$\ 0.30 & 0.93 & 0.94 & 1.07 & 2 \\
  3656121 & 13.64\ $\pm$\ 0.28 & 0.00019\ $\pm$\ 0.00069 & 6212\ $\pm$\ 98 & 4.38\  $\pm$\ 0.30 & 1.09 & 1.21 & 1.17 & 2 \\
  3832474 & 16.19\ $\pm$\ 1.77 & 0.00388\ $\pm$\ 0.00363 & (...) & (...) & 0.99 & 0.95 & 8.94 & 1 \\
  3967760 & 15.11\ $\pm$\ 1.45 & 0.00294\ $\pm$\ 0.00297 & 5372\ $\pm$\ 113 & 4.48\ $\pm$\ 0.30 & 0.95 & 0.94 & 6.18 & 2 \\
  \enddata 
  \tablenotetext{a}{~Full table available in the online version. 
  A portion is shown here for guidance regarding its form and content.}
  \tablenotetext{b}{~Uncertainties from Monte Carlos Markov chains estimation, see \S~\ref{Mcmc}.}
  \tablenotetext{c}{~Values from \citet{pinsonneault-12}.}
  \tablenotetext{d}{~Values from KIC.}
  \tablenotetext{e}{~Values from KCP Public Archive.}
  \tablenotetext{f}{~Level of period reliability (1 $=$ higher, 2 $=$ lower).}  
  \label{Table}
\end{deluxetable*}

\subsection{Pre--selection of KCP} \label{PreSelKCP}

For our main sample, harmonic fits of the LCs were computed to estimate their
variability amplitudes, as described in \citet{demedeiros-13}, and the fit residual was
assumed to be the LC noise. Thus, the peak--to--peak amplitude over the standard
deviation of the LC noise was defined as the variability signal--to--noise $S/N$ ratio
\citep[see][]{demedeiros-13}.  Simulations from \citet{demedeiros-13} \citep[see Fig.~2
of][]{demedeiros-13} suggest that a reasonable pre--selection can be obtained by using a
cutoff of $S/N > 1.0$, when the sample is sufficiently small to be visually inspected. 

After this automatic pre--selection, the resulting subsample was visually
inspected by applying the list of criteria described by \citet{demedeiros-13}, which serve to define
the photometric signature of rotational modulation caused by dynamic star spots
(namely semi--sinusoidal variation\footnote{In short, semi--sinusoidal variability is
characterized by some asymmetry of the maximum and minimum fluxes with respect to the
average flux over time, somewhat irregular long--term amplitude variations, and semi--regular
multi--sinusoidal short--term flux variations with a typical amplitude of
$\lesssim$~0.5~mag and typical period of $\gtrsim$~0.3~d. This description was developed based on the
dynamic behavior of star spots, as observed on the Sun.}). In addition,
\citet{bravo-14} performed an unprecedented comparison of different variability
signatures in wavelet maps, including rotational modulation.  This approach was also
considered in our visual inspection procedure for better identification photometric variability induced
by spots.  As another selection step, we considered the fact that in certain LCs (approximately 5\%
of a sample) the rotation period may correspond to some multiple of the main periodogram
peak, instead of the main peak itself, as explained in \citet{demedeiros-13}.  This phenomenon
may be observed simultaneously in the Lomb--Scargle and wavelet periodograms for the same LC.
\citet{demedeiros-13} proposed that in such cases, the phase diagram of the LC typically exhibit 
two sub--cycles of different depths, which compose a full cycle.  For the sake of
simplicity, we excluded such cases through visual inspection. 

From the 408 KCP stars considered in our parent sample, only LCs with $S/N >1.0$ 
that exhibited clear semi--sinusoidal signatures were selected. This selection resulted in a final
sample of 131 stars.  Finally, we flagged LCs for which the periods derived using the Lomb--Scargle 
and wavelet methods agreed with $\sim$10\%. These were used as an high confidence subsample for evaluation of the results. 
This subsample of high confidence periods comprised 67 of the 131 stars (see Table~\ref{Table}).

\subsection{Pre-selection of KOIs} \label{PreSelKOIs}

Because the KOIs were used as a comparison sample, a stricter automatic
pre--selection was applied. Only LCs with $S/N > 2.0$ for which the periods derived using 
the Lomb--Scargle and wavelet methods (agreed within $\sim$10\% discrepancy) were selected.
This automatically pre--selected sample was then visually inspected in the same manner as for the
KCP host stars.  In this step we selected only LCs with transits fainter than the peak--to--peak
amplitude of the semi--sinusoidal variability. Of the $3\!,798$ KOI stars considered
in our parent sample, 409 objects were selected for comparison with our KCP subsample.

\subsection{Angular momentum} \label{AngularM}

For all of the stars in our final sample the stellar angular momentum was estimated
following the procedure proposed by \citet{alves-10}. When possible, we also
estimated the total angular momentum, i.e., the angular momentum of the host stars
plus the contribution of its planets, using the following simple central force model:
\begin{equation}
  J_{star-pl}=\mu \ \sqrt{G(M_{star}+M_{pl})a(1-e^{2})}
\end{equation}
where $\mu$ is the reduced mass, $G$ is the gravitational constant, $M_{star}$ is the mass of the
host star, $M_{pl}$ is the total mass of all planets in the system, $a$ is the semi--major axis 
of the orbit, and $e$ is the eccentricity of the orbit.  The uncertainty in the angular momentum was 
calculated via error propagation, using the uncertainties in the stellar mass and radius provided in the NASA
Exoplanet Archive\footnote{\anchor\url{http://exoplanetarchive.ipac.caltech.edu/}}.

\subsection{Error estimation} \label{Mcmc}

In principle, we can simply derive \period from the output of the Lomb--Scargle periodograms and the 
wavelet method, and then use these results to extract other features (i.e., the amplitude and phase) 
through least squares fitting. Some authors \citep[e.g.,][]{walkowicz-13} suggest estimating the period 
uncertainty from the periodogram, whereas others \citep[e.g.,][]{kovacs-81,horne-86} suggest using a 
combination of features from the light curve data, a fit of the rotational period, and the periodogram.

We propose a different approach that is based on Bayesian analysis, which provides a natural framework 
for the estimation of both a model and its uncertainty. The method proceeds as follows: the prior probability 
distribution function provides the method with knowledge of the parameters and their uncertainties before the 
observational data are incorporated. The likelihood function provides the information regarding the data itself. 
The posterior probability distribution function is then constructed from these two inputs.

We introduced the parameters from the period fit ($P_{\rm rot}$, amplitude, phase) into the prior probability 
distribution function. We also introduced the possibility that the calculated \period may be twice or half the 
real rotational period \citep[see \S~2.2.2 in][]{demedeiros-13}. The potential errors in the flux, which were 
calculated from the \emph{Kepler} Flexible Image Transport System (FITS) files and nearest--neighbor flux dispersion 
in the LCs \citep[see Eq.~1 in][]{demedeiros-13}, were introduced into {the} likelihood function.

Markov chain {Monte Carlo} sampling was used to implements a modification of the Metropolis--Hastings algorithm 
developed by \citet{goodman-10}. This approach offers the advantage of requiring only simple hand--tuning and thus, 
producing samples more rapidly than traditional algorithms. To achieve this implementation we followed the procedure 
proposed by \citet{foreman-13}. For each LC, $1\!,000$ iterations were performed to generate a virtual sample, 
based on the information obtained from the Lomb--Scargle and wavelet analyses along with our degree of uncertainty. 
Thus the outcome distribution was rooted in our primary calculations. The median of the Bayesian \period distribution 
differed slightly from our \period values (an average difference of 0.16\%), thus its standard deviation reflected the 
period uncertainty of the LC fit (see Table~\ref{Table}).

\section{Results and Discussion} \label{Results}

\begin{figure}[tb] 
  \plotone{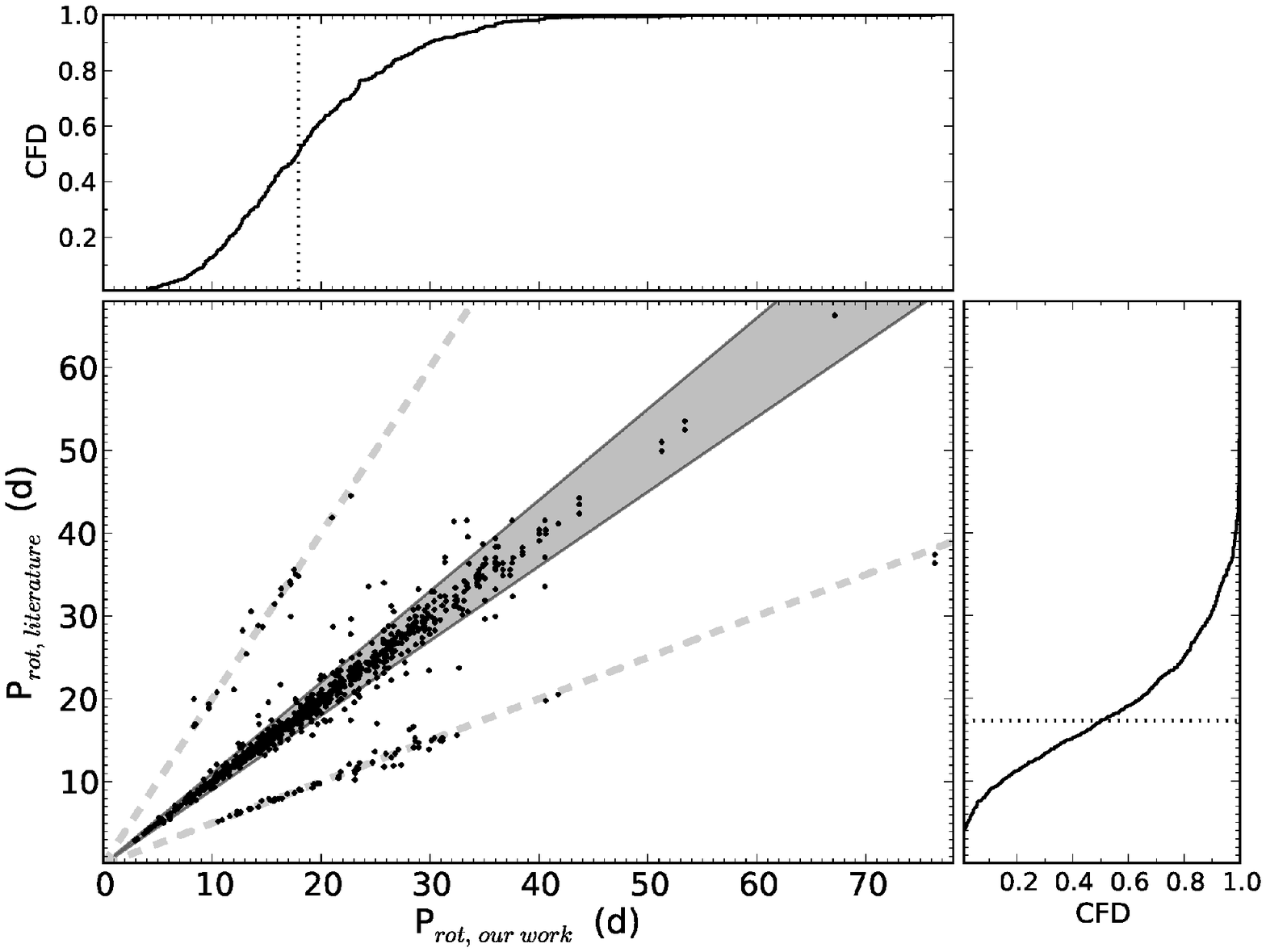}
  \caption{
      Comparison of the rotation periods computed in our work with those from the literature
      \citep{mcquillan-13a,mcquillan-13b,mcquillan-14,nielsen-13,reinhold-13,walkowicz-13}. The grey 
      area indicates the region that represents a discrepancy of 10\% or less, and the dashed 
      lines represents 2 \period aliases. Cumulative frequency distributions (CFDs) are presented 
      for the horizontal (panel at the top) and vertical (panel to the right) axes, in which the 
      dotted lines mark the median values (both approximately 18~d). 
      }
  \label{comparison}
\end{figure}

\begin{figure}[h!tb] 
  \plotone{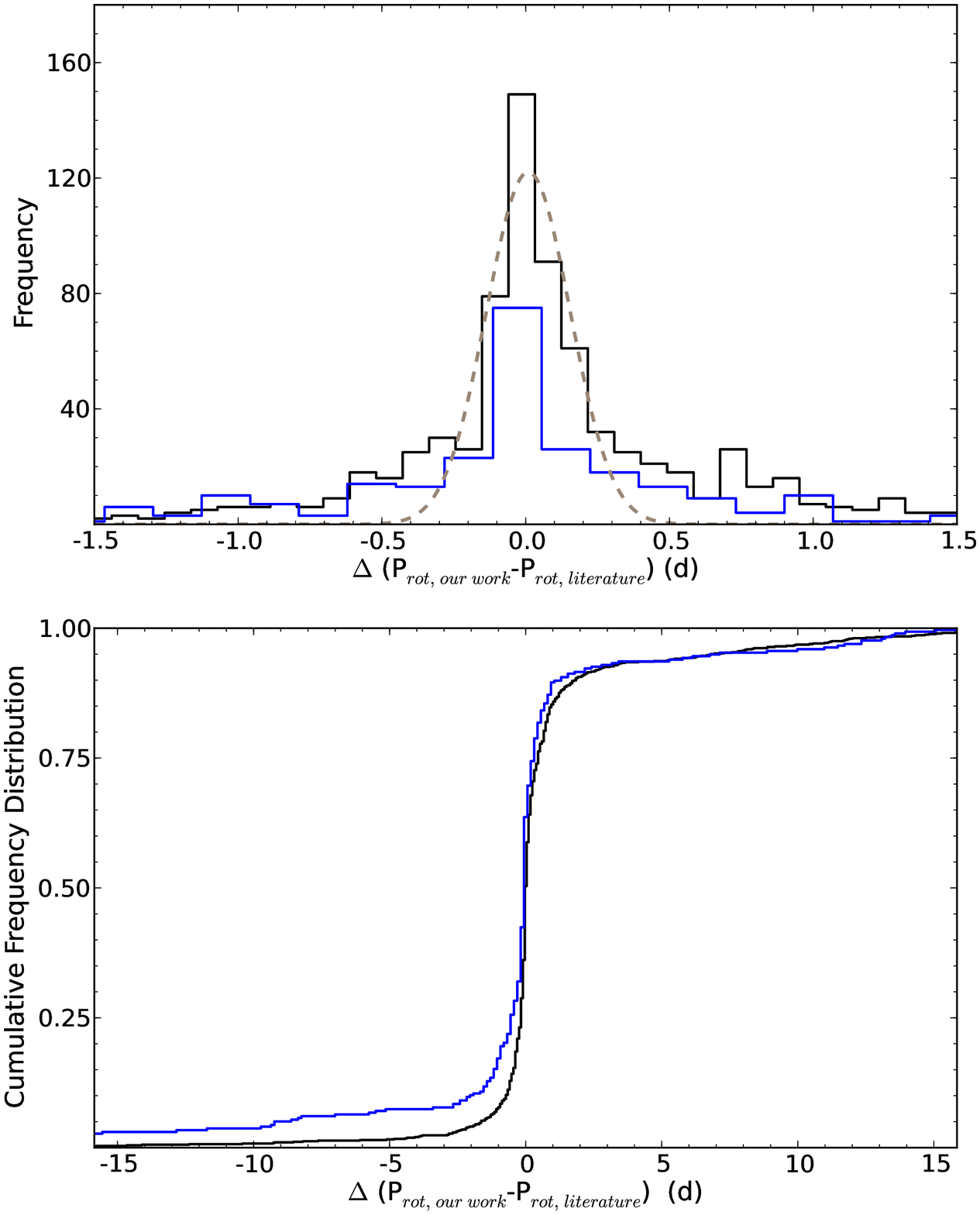}
  \caption{
      \textit{Upper panel}: Histogram of the difference in the periods from our work and those 
      reported in the literature, from Fig.~\ref{comparison}. The solid black histogram 
      represents KOIs and the solid blue histogram represents KCP host 
      stars. A Gaussian fit with the sum of the KOIs and the KCP host stars is depicted by 
      the dashed gray curve (mean of 0, FWHM of 0.32, and standard deviation of 0.14) to contrast 
      the wings of the black and blue histograms.
      \textit{Lower panel}: CFDs for the difference in the \period values determined in our work 
      and those presented in the literature for KOIs and KCP host stars. The color coding is the 
      same as in the upper panel. 
      Although the Kolmogorov--Smirnov and Anderson--Darling tests suggest different 
      parent distributions, we verified this is a bias effect (see text).
      }
  \label{histo}
\end{figure}

We identified a total of 477 stars {that were} in common with previous works related to our study \citep{mcquillan-13a,mcquillan-13b,mcquillan-14,nielsen-13,walkowicz-13,reinhold-13}.  For those stars, a comparison between our periods and those previously published is presented in Fig.~\ref{comparison}. According to this figure, our period calculations are in strong agreement with the literature values.  
This agreement can also be observed in the upper panel of Fig.~\ref{histo}, in which a Gaussian fit to the histogram of $P_{\rm rot,\ our\ work}-P_{\rm rot,\ literature}$ yields a full width at half maximum (FWHM) of 0.32~d, a value that is comparable to the lowest \period uncertainties in our work.
From Fig.~\ref{comparison}, the existence of harmonics at $P_{\rm rot,\ literature}=2P_{\rm rot,\ our\ work}$ and $P_{\rm rot,\ literature}=P_{\rm rot,\ our\ work}/2$ can be clearly observed. It is difficult to unequivocally determine \period in such cases if the $S/N$ value is low, but because we set a sufficiently $S/N$ threshold \citep[according to \S~2.2.2 of][]{demedeiros-13} and performed visual inspection of all LCs, we have confidence in our results. A more thorough comparison that is focused only on these objects is underway.
The lower panel of Fig.~\ref{histo} presents the CFDs of the difference in our \period values and those from the literature, for both the KCP host stars and KOIs. It can be inferred that the harmonic at $P_{\rm rot,\ literature}=2P_{\rm rot,\ our\ work}$ can be predominantly attributed to the KCP host stars, the sample on which we focused more attention. In addition, the results of the Kolmogorov--Smirnov test (distance of 0.117 and probability of $4.4 \times 10^{-3}$) and the Anderson--Darling test (asymptotic probability of $9.5 \times 10^{-5}$) performed using the \texttt{R} software package \citep{R,AD} strongly indicate that the two distributions are drawn from different parent distributions. 

Based on Fig.~\ref{histo}, the KCP sample would have shorter period values in the present work
than in the literature, in average, thus being more populated at negative values
of the abscissa than the KOI sample. However, this apparent discrepancy is actually a bias
produced by a trend of computing a set of \period values in the present work that are an half of those
provided in the literature. This is an effect of the distinction between the procedures for measuring
\period and avoiding their aliases (see \S~\ref{PreSelKCP}). In fact, the large data sets from the 
literature were typically developed from semi--automatic methods. A semi--automatic method, which was 
also performed in the present work to build the KOI sample, is subject to produce false positives, 
especially due to aliases. In contrast, the KCP sample was obtained with a considerably more careful 
treatment, inspection, and selection of the LCs and should have less aliases than the KOI sample.
To avoid such a bias, we can remove the doubtful cases by considering that the shaded area in 
Fig.~\ref{comparison} represent the most confident period measurements for both this work and 
literature. Indeed, if we select only the sample belonging to the shaded area in Fig.~\ref{comparison}, 
the KOI and KCP distributions become highly similar: Kolmogorov--Smirnov test yields a distance of 0.0214 
and a probability of $0.97$; Anderson--Darling test yields an asymptotic probability of $0.99$.

For the total of 477 publicly available periods that are in common with our sample, this work offers a cross--validation between our calculated periods and the previously published values. In addition, we provide stellar rotation periods of 63 stars for which values were not previously reported in the literature.  

Wavelet maps provide important diagnostics for the behavior of starspot dynamics \citep{defreitas-10,bonomo-12,sello-13}. The main period, as observed in the wavelet maps and power spectra, is typically the rotation period, as is commonly assumed in the literature \citep[e.g.,][]{garcia-14,mcquillan-13a,mcquillan-13b,mcquillan-14,nielsen-13,walkowicz-13,reinhold-13}, except for approximately 5\% of a sample \citep[][see also \S~\ref{PreSelKCP}]{demedeiros-13}. The second period is an harmonic that may be an effect of the superposition of active regions and their dynamics.  

To test our method, we analyzed the hourly total solar irradiance (TSI) and spectral solar irradiance (SSI) variation in data from the Variability of Solar Irradiance and Gravity Oscillations \citep[VIRGO,][]{virgo} instrument on Solar and Heliospheric Observatory (SOHO) satellite\footnote{ftp.pmodwrc.ch/pub/data/irradiance/virgo/}, as in \citet{lanza-04}.  These authors verified, based on {the} wavelet and Lomb--Scargle methods, that the rotation period of the Sun can be effectively estimated from TSI and SSI data in the solar minimum phase, except when the solar photometer (SPM) is centered at 500~nm of farther from the solar minimum. In the case of this exception, the rotational modulation is masked by the typical relaxation time of the active regions, which is approximately 5--7~d \citep{lanza-04}. When our method was applied to VIRGO SSI \textit{green} data (centered at 500~nm), the wavelet and Lomb--Scargle methods yielded results that were similar to those of \citet{lanza-04}. In this manner, we detected a main period of approximately 7.3~d (6.53~d for the wavelet method and 8.11~d for the Lomb--Scargle method) for a very particular case (ie. at 500~nm and farther from the solar minimum, cycle 23). Considering the \emph{Kepler} Response Function (which is centered at 600~nm, with $\delta=^{+300}_{-200}$~nm), which is compatible with the VIRGO SSI \textit{green} data, we expect to correctly detect the rotational modulation of Sun--like stars.  

\begin{figure}[t!b] 
  \centering 
  \plotone{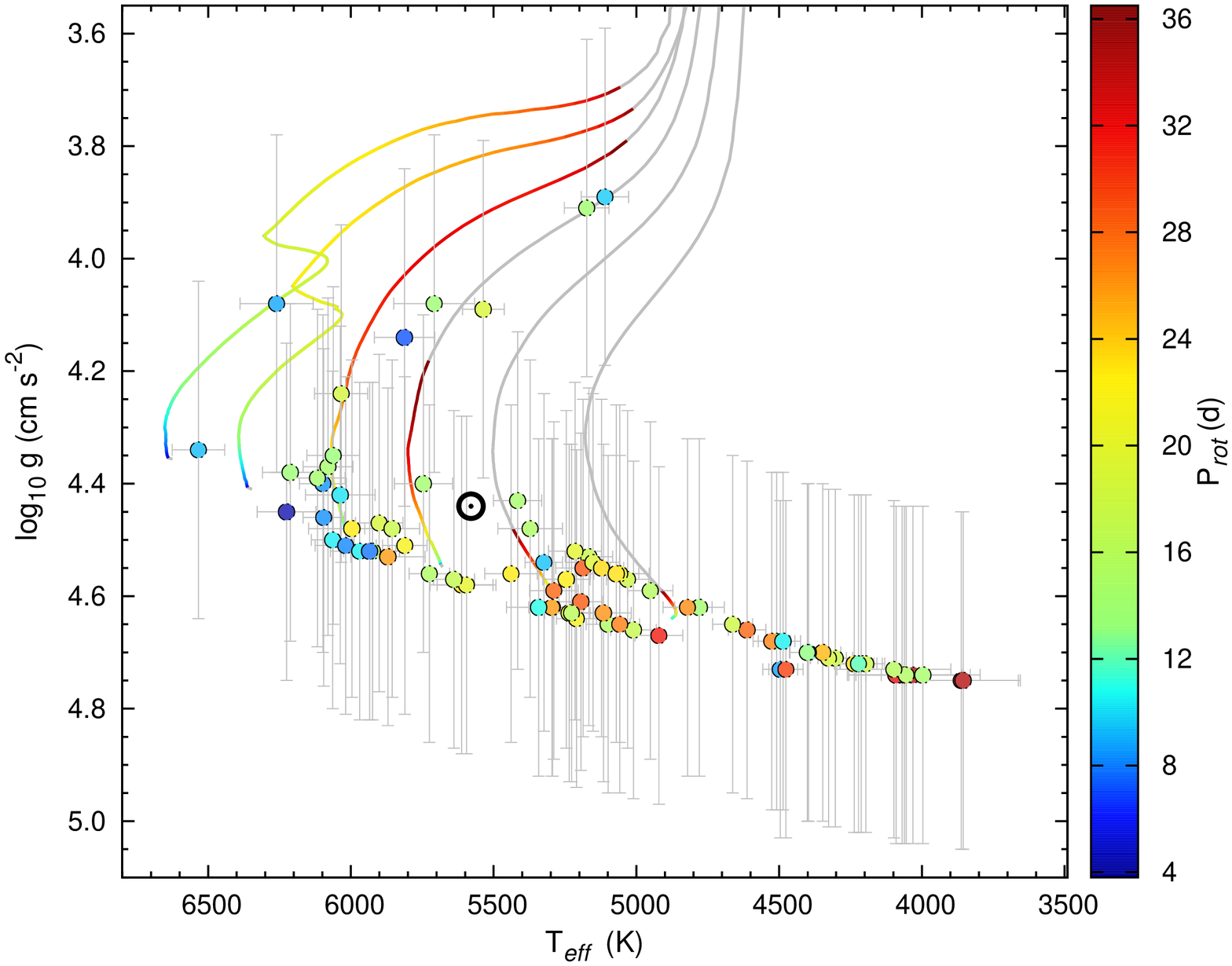}
  \plotone{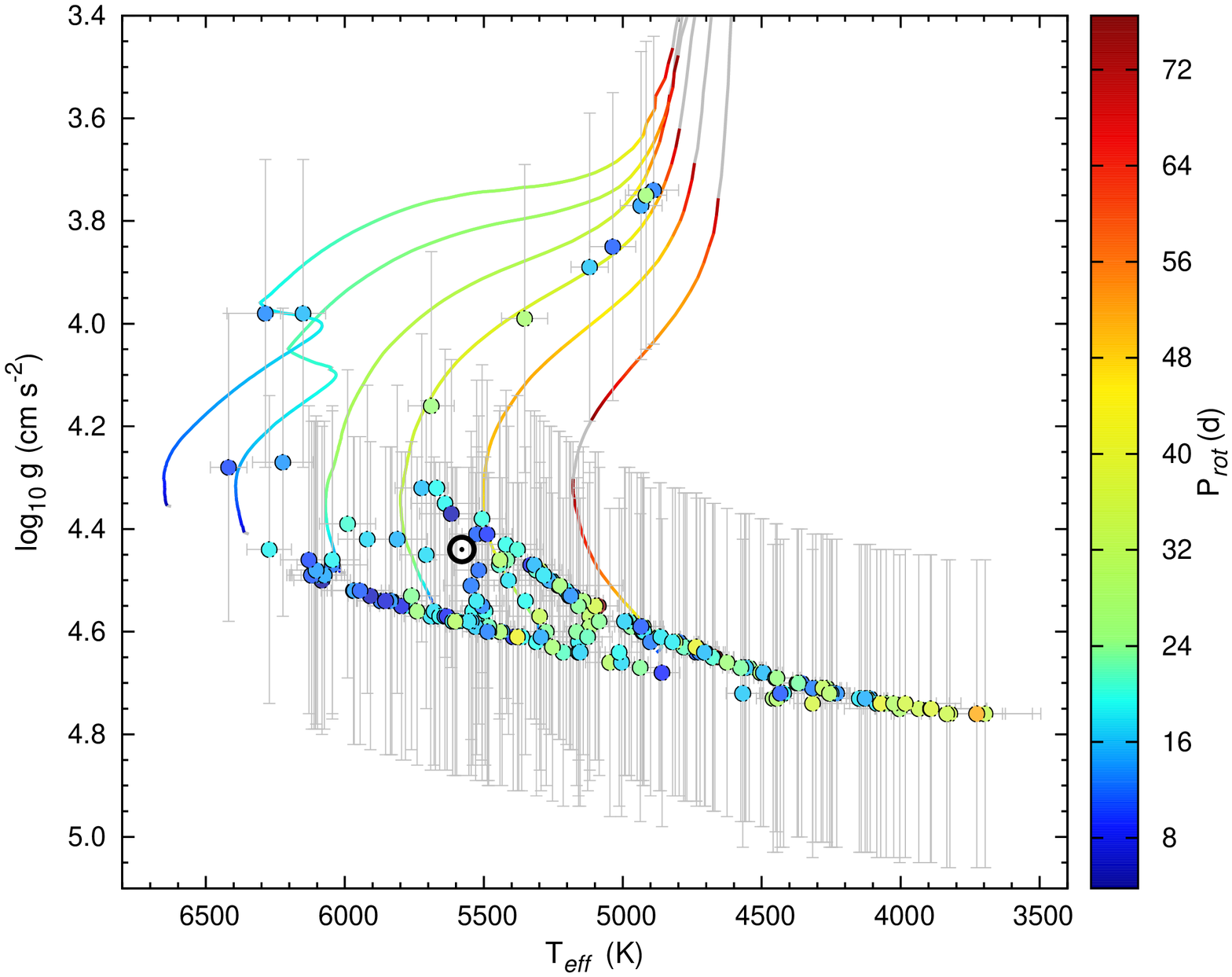}
  \caption{HR diagrams for the KCP (upper panel) and KOI (bottom panel) samples,
          overplotted with {evolutionary tracks that account for rotation from} \citet{ekstrom-12}  From
          left to right, the tracks displayed correspond to initial masses of: 1.35, 1.25, 1.1,
          1.0, 0.9 and 0.8~$M_{\sun}$. Their colors represent the theoretical $P_{\rm rot}$ values,
          as indicated by the color scale on the right (in units of days), and values outside of 
          this range appear in gray.  Filled circles represent our data, for which the colors correspond to the
          same color scale used for the tracks. The Sun is represented by its usual symbol.}
  \label{loggteff}
\end{figure}

Figure~\ref{loggteff} displays the distribution of rotation period in the HR diagram for 85 KCP host stars (upper panel), and 193 KOI stars (bottom panel), where \teff and \logg were obtained from \citet{pinsonneault-12} and the \emph{Kepler} Input Catalog (KIC), respectively. Theoretical tracks for solar metallicity ($Z=0.014$) from \citet{ekstrom-12} are included.  These tracks span a range of 0.8 to 120~M$_{\sun}$, {beginning with} zero--age main sequence (ZAMS) stars, with an initial rotational velocity of 0.4 times the critical velocity, and ending with the early asymptotic giant branch (AGB). The model accounts for the magnetic braking law \citep[see][for a nonextensive approach]{defreitas-13a}, and sets abundances that are compatible with the solar values. As revealed by Fig.~\ref{loggteff}, the present stellar sample is composed of stars with masses that range from those of low--mass M--type stars (namely, the coolest stars in the sample with a minimum mass of 0.48~M$_{\sun}$), up to 1.53~M$_{\sun}$. These mass estimations were obtained directly from KCP\footnote{\anchor\url{http://kepler.nasa.gov/Mission/discoveries/}} when available, and from the NASA Exoplanet Archive otherwise. The observed range of rotation periods agrees well with the position of the stars in the HR diagram. For example, there is an important number of low--mass stars, namely those with \teff values of less than approximately $4\!,870$ K, {for which the} period values agree with the distribution of rotation period for M stars previously reported by other authors \citep[e.g.,][]{irwin-11}.

One interesting aspect of the present analysis concerns the stars with \teff and \logg parameters that are similar to those of the Sun. We have identified 5 KCP and 12 KOI stars with $T_{\rm eff}$ {values within 100~K of} $T_{\rm eff\, \sun}$ and $log\, g$ values within 0.10~dex of  $log\, g_{\sun}$, which, within the uncertainties, are consistent with the solar values. These stars exhibit rotation period that range from 7 to 26~d. Note that the rotation period of the Sun ranges from 24.47 days at the equator to 33.5 days at the poles and is equal to 26.09 days on average \citep{lanza-03}. The stars that have \teff and \logg values that are similar to the solar values are the KCP stars KIC~3541946, KIC~5351250, KIC~9455556, KIC~11389771, KIC~11709244, and the KOI stars KIC~5360920, KIC~6226290, KIC~6392727, KIC~6599919, KIC~6869184, KIC~7382313, KIC~7870032, KIC~8676148, KIC~9839821, KIC~10811496, KIC~11565544, and KIC~12644822 . Of particular interest are KIC~9455556, KIC~11565544, and KIC~12644822, which have rotation periods that are similar to the Sun\textquoteright s, namely 25.78~d, 24.93~d, and 23.51~d, respectively.

The 523 remaining sources exhibit super-- and sub--solar rotation, with periods that range from 2.05 to 88.94~d (see Table \ref{Table}). The period distribution of our sample agrees with that of \citet{reinhold-13}, with a maximum of approximately 10--20~d. 

\begin{figure}[t!] 
  \centering 
  \plotone{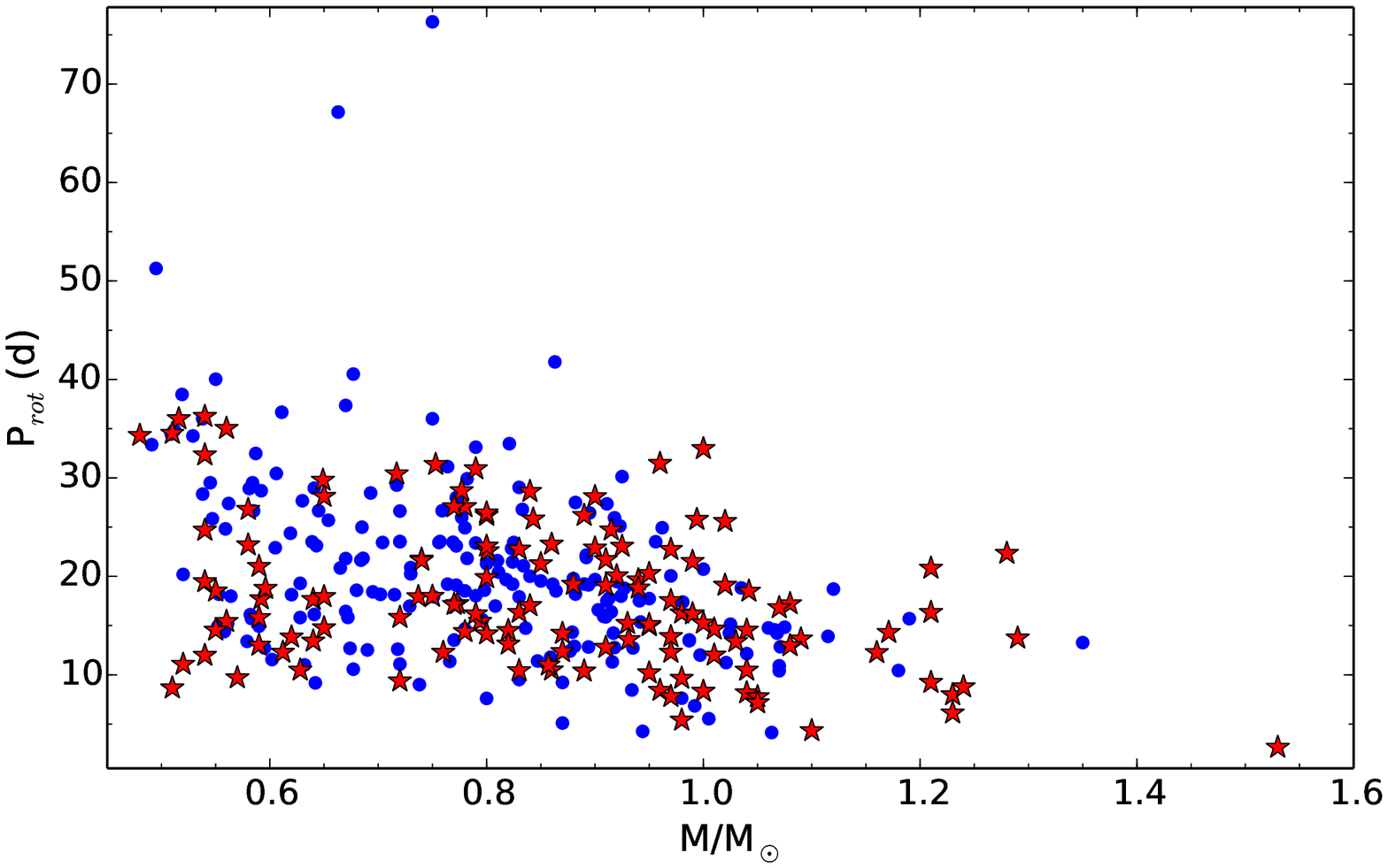}
  \caption{Period--mass diagram for the KCP (red stars) and
          KOI (blue circles) samples.}
  \label{permass}
\end{figure}

When the stellar masses are compared with the corresponding \period values, as in Fig.~\ref{permass}, the expected decay trend is observed: faster rotators have higher masses (engulfment can be invoked for the rapid rotators with signatures of close planets). However, because of the range of masses that we investigated we did not observe the change in slope that was detected by \citet{mcquillan-13a}.

Another notable case is that of the star KIC~12735740, whose spectroscopic parameters are given by \citet{wang-13}. This target exhibits values of temperature, surface gravity, mass, and radius that are similar to the solar values, but it has a lower P$_{\rm rot}$ (19.64~$\pm$~0.61 d), similar to the case of $\epsilon$ Eridani \citep{metcalfe-13}. There is a notable spectral similarity between KIC~12735740 and the Sun in the \ion{Ca}{2} H region \citep[see Fig.~3 {of}][]{wang-13}, which leads those authors to conclude that this star is chromospherically inactive.  In support of this conclusion, the star has a very slow projected rotation velocity, with $v\ \sin i$ \citep[1.43~$\pm$~0.78~km~s$^{-1}$,][]{wang-13}, which is similar to the solar value \citep[1.6~$\pm$~0.3~km~s$^{-1}$,][]{pavlenko-12}.
The apparent discordance that arises from KIC~12735740 exhibiting a smaller \period than that of the Sun but a similar $v\ \sin i$ value, depends on the values of the microturbulence and macroturbulence, along with the inclination angle. This angle can contribute significantly to the projected velocity. Therefore a strict comparison (which is plausible only in the case of homoscedasticity) is difficult to perform. To reinforce this claim, it should be noted that $\epsilon$ Eridani has a period of 11.20~d \citep{croll-06} and a $v\ \sin i$ value of 2.45~km~s$^{-1}$ according to \cite{fischer-05}, but this value decreases to 1.7~km~s$^{-1}$ in the photometric--based measure of \cite{croll-06} (see \S~{5.2} in the cited study). Its temperature of $5\!,156$~K \citep{clem-04} is slightly sub--solar, nevertheless, its surface gravity \citep[4.57~dex,][]{clem-04} is consistent with the solar value. Other relevant parameters of $\epsilon$ Eridani include its radius and mass, which are similar to the solar values \citep[0.82 \msun and 0.74~$R_{\sun}$][]{baines-12}, and its younger age of 850~Myr \citep{difolco-04}. Therefore, care must be taken in any direct comparison that is based solely on $v\ \sin\ i$.

\begin{figure}[tp] 
  \centering 
  \plotone{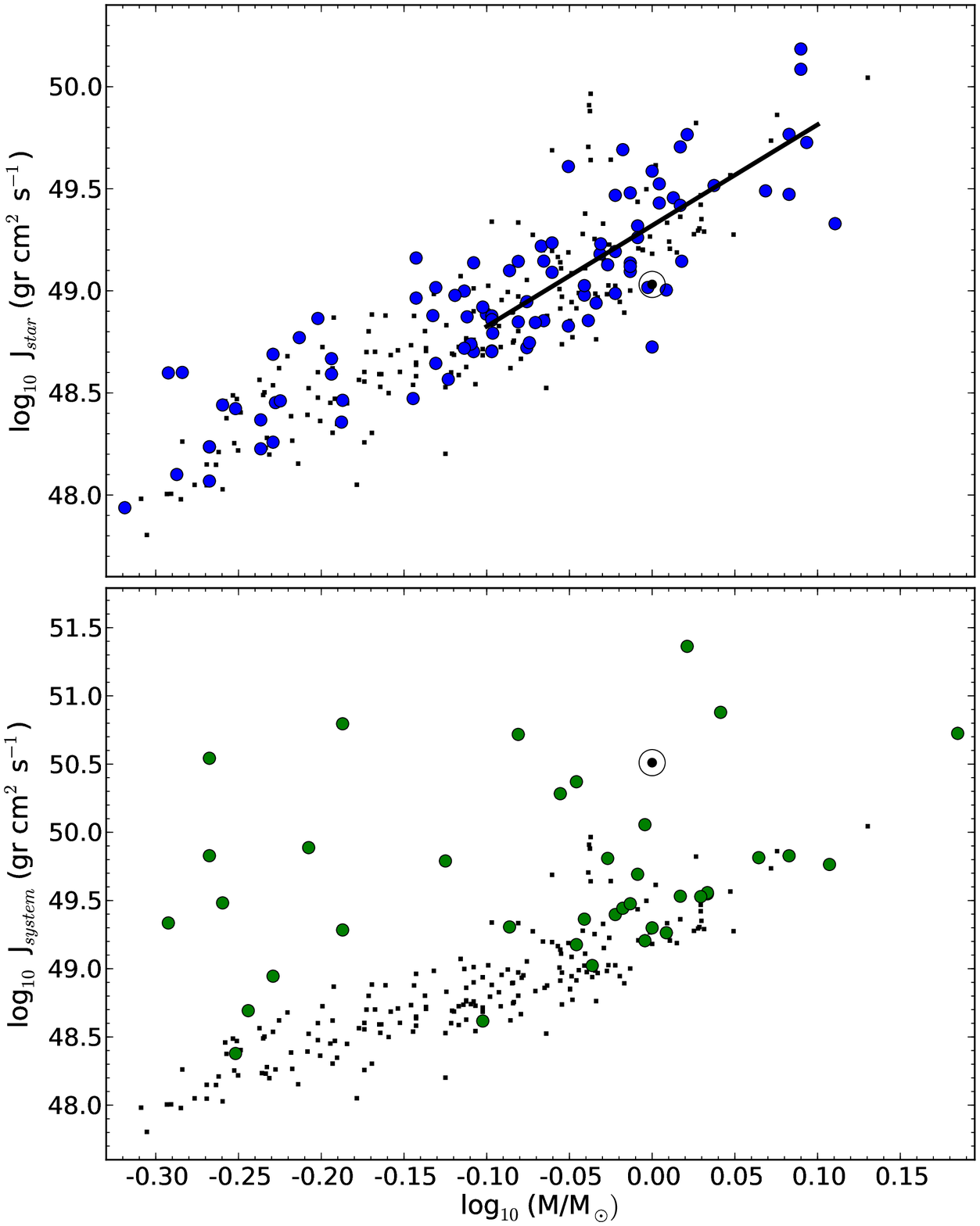}
  \includegraphics[scale=0.325]{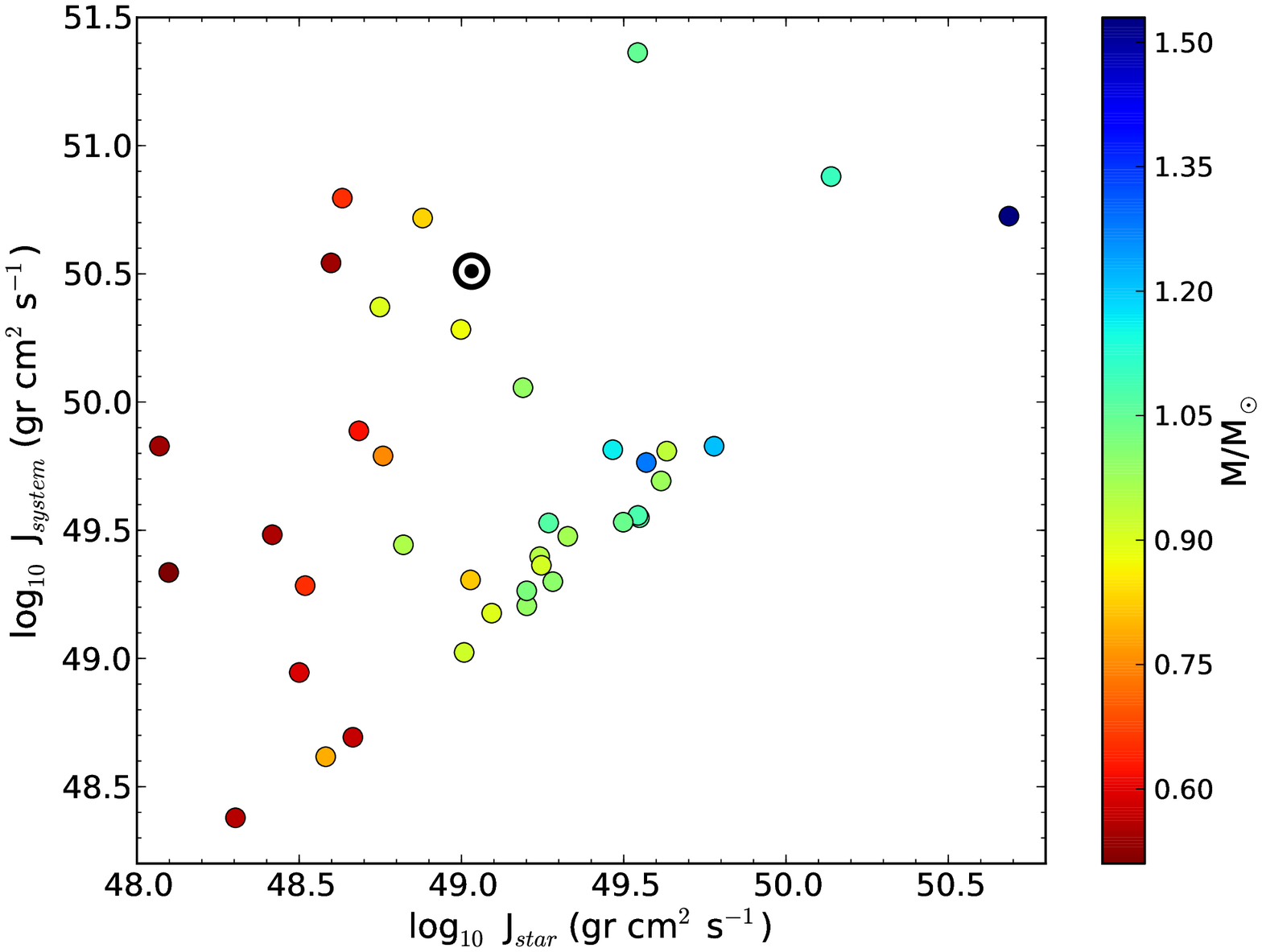}
  \caption{\textit{Upper panel}: stellar angular momentum as a function of
          stellar mass for our final sample of 131 KCP host stars (blue circles), and a
          sub--sample of 193 KOI stars (black squares). 
          \textit{Middle panel}: total angular momentum of the star--planet systems for 38 (of the total of 131) 
          KCP stars (green circles), overplotted with
          the stellar angular momentum of the KOI sample (black squares). 
          \textit{Lower panel}: stellar angular momentum versus total angular momentum of the star--planet 
          systems where the color indicates the mass of the host star. A clear dependence of the 
          mass on $J_{star}$ is observed. In all panels, the Sun is represented by its usual symbol.} 
  \label{j_mass}
\end{figure}

One of the most notable characteristics of the solar system is its large amount of angular momentum, which is largely associated with Jupiter, which possesses at least two orders of magnitude more orbital angular momentum than the spin angular momentum of the Sun itself.  In this context, we also analyzed the stellar angular momentum, which was obtained using the relation $J_{star} \propto (M_{star}/$M$_{\sun})^{\alpha}$, (this is the stellar contribution only, with no planetary contribution) of the stars that compose the present sample, to examine how their spins compare with that of the Sun. For the $\alpha$ uncertainty we performed a Markov chain Monte Carlo analysis, in which, Kraft\textquoteright~{s} relation \citep{kraft-67} was employed as the fit to the model and fluctuations compatible with stellar angular momentum uncertainties were then applied to the fit. A total of $10\!,000$ iterations were performed and the uncertainty was defined as in \S~\ref{Mcmc}.
Figure~\ref{j_mass} presents the distribution of the stellar angular momentum $J$ as a function of stellar mass for our sample of \emph{Kepler} stars with planets, for which the $J$ values were computed following the procedure proposed by \citet{alves-10}. The Sun is represented by its usual symbol. The solid line illustrates the best fit of Kraft\textquoteright s relation $J_{star} \propto (M_{star}/$M$_{\sun})^{\alpha}$ ($\alpha$=$4.9 \pm 1.4$) \citep{kraft-67} with our star sample, where the exponent $\alpha$ of the power law was treated as a free parameter. Clearly, the distribution of the stellar angular momentum for stars with masses of greater than approximately 0.80~\msun follows, at least qualitatively, the law $J \propto (M/$M$_{\sun})^{\alpha}$, as previously demonstrated by \citet{alves-10} for stars with planets detected using the Doppler method. The mass range of applicability of this relation is as follows: its lower end is given by the mass limit of F-- and G--type stars (approximately 0.8~M$_{\sun}$). Conversely, the upper mass limit (approximately 1.25~M$_{\sun}$) was set based on \citet{kawaler-87}, according to whom such stars possess thin convection zones. Above this mass regime the relation between angular momentum and mass obeys a different power law \citep[the high--mass relation of \S~{3} in][]{kawaler-87}, because of the emergence of a radiative outer zone; consequently, the influence of stellar convective transport in stellar envelopes rapidly decreases its influence on the relation between angular momentum and mass for high--mass stars.

Of the total working sample of 131 KCP host stars, the required parameters for the computation of the total angular momentum are present in the literature for only 38 stars. The bottom panel of Fig.~\ref{j_mass} displays the distribution of the total angular momentum of the star--planet systems of these stars as a function of stellar mass. The Sun symbol in this figure corresponds to the total angular momentum of the solar system. Interestingly, although most of the planet hosting \emph{Kepler} stars exhibit a trend of excess in stellar angular momentum compared with that of the Sun, as observed in the upper panel of Fig.~\ref{j_mass}, the angular momentum of the star--planet systems exhibits a somewhat different behavior, with a number of KCP host stars exhibiting total angular momenta that are comparable to that of the solar system and other KCP host stars exhibiting a deficit in total angular momentum, compared with the Sun.

In addition, \citet{alves-10} demonstrated that stars that host more massive planets tend to have higher angular momenta, whereas stars that host less massive planets tend to have lower angular momenta. We tested these findings, namely, the results presented in Fig.~9 of \cite{alves-10}, using our sample and, in contrast with the results of the previous authors, no clear dependence of the stellar angular momentum on the planetary mass was observed.

There are at least two possible explanations for these results: (i)~the relatively low-number of stars in the KCP sample may hinder a proper statistical treatment of the results, and (ii)~an important number of KCP sources may host undetected planets, thus biasing the results. The latter explanation suggests the potential for the detection of new planets in the KCP sample.  However, another possible explanation for the discrepancy between the present finding and that of \citet{alves-10} resides in the procedure used for the detection of the planets. The sample selected by \citet{alves-10} was based on radial velocities, whereas the present sample was chosen based on eclipse data, in which the handicap of detecting planets closer to the stellar host is not present because of the typical observation window of the \emph{Kepler} satellite.

\section{Conclusions} \label{Conclus}

We analyzed photometric variations in the current sample of \emph{Kepler} Planetary Host
Stars and determined \period for a final sample of 131 sources.  These periods were
calculated independently via the Lomb--Scargle and wavelet methods, thereby yielding a reliable
\period determination.  As remarked in \S~\ref{KeplerData}, this unified method allowed us
to consistently distinguish \period from other manifestations of magnetic activity, even when
the associated amplitude variations were of the same order of magnitude.

Our \period values are consistent with those previously reported in the literature by
\citet{mcquillan-13a,mcquillan-13b,mcquillan-14}, \citet{nielsen-13},
\citet{walkowicz-13}, and \citet{reinhold-13}. They are also in strong agreement with the
theoretical predictions of \citet{ekstrom-12}.  In particular, the agreement between the tracks from
\citet{ekstrom-12} and our data is clear and provides good experimental support for stellar rotation 
theory, despite possible different initial conditions between the models and observed stars. 

The present analyses has also revealed an interesting group of 5 KCP and 12 KOI stars with $T_{\rm eff}$ and $log\, g$ 
values similar to those of the Sun and rotation period ranging from 7 to 26 days. Of particular interest are the 
stars KIC~9455556, KIC~11565544, and KIC~12644822, rotating with periods similar to the Sun\textquoteright s values, 
namely 25.78~d, 24.93~d, and 23.51~d, respectively.

Finally, the stellar angular {momenta} of our subsample of 131 KCP and 193 KOI stars follow Kraft\textquoteright~{s} 
relation, thus offering an important generalization of this law for a particular sample with known planets and
photometric \period measurements. Despite this result, no relationship between stellar angular momentum and planetary 
mass was found, in contrast with the results of \citet{alves-10}.

Open questions remain because robust statistics from a larger planetary host sample are required. 
For example, to what extent do low \period values reflect a bias in detectability? How does planetary systems affect 
the $P_{\rm rot}$ value of the central star? A larger sample will also help elucidate the behavior
of angular momentum with respect to planetary mass, which is a crucial ingredient in the 
modeling of star--companion systems. Special attention must be payed to the KOI and KIC stars presenting 
Sun\textquoteright s rotation rate and $T_{\rm eff}$ and $log\, g$ solar values, for which a solid spectroscopic 
study could show how close of the present day Sun are their evolutionary stages.

  \acknowledgments F.P.-Ch. acknowledges his wife, for her warm and continuous support beyond the realm of science.
  We thank the anonymous referee for suggestions that helped improve our work.
  This paper includes data collected by the \emph{Kepler} mission. Funding for the {\em Kepler} mission is provided by the NASA Science Mission directorate.  All data presented in this paper were obtained from the Mikulski Archive for Space Telescopes (MAST). STScI is operated by the Association of Universities for Research in Astronomy, Inc., under NASA contract NAS5-26555. Support for MAST for non--HST data is provided by the NASA Office of Space Science via grant NNX13AC07G and by other grants and contracts. This paper utilized the NASA Exoplanet Archive, which is operated by the California Institute of Technology under contract with NASA under the Exoplanet Exploration Program.
  We acknowledge the use of the CasJobs facilities, and the JHU/SDSS team for this powerful tool.
  The research activity of the Stellar Board of the Federal University of Rio Grande do Norte (UFRN) is supported by continuing grants from the Brazilian agencies CNPq and FAPERN. We also acknowledge continuing financial support from INCT INEspa\c{c}o/CNPq/MCT.
  F.P.-Ch. acknowledges a doctoral fellowship {from} CNPq. J.P.B. acknowledges doctoral fellowships from the Brazilian agency CAPES. S.A. acknowledges a post-doctoral fellowship from CAPES (BEX-2077140) and the support of the Iniciativa Cient\'{i}fica Milenio through grant IC120009, which was awarded to the Millennium Institute of Astrophysics (MAS). I.C.L. acknowledges a post-doctoral fellowship from PNPD/CNPq. C.E.F.L. acknowledges a post-doctoral fellowship from PDJ/CNPq.
  F.P.-Ch. and M.C. acknowledge additional support from the Programa Iniciativa Cient\'{i}fica Milenio of the Ministry for the Economy, Development, and Tourism through grant P07-021-F, which was awarded to the Millennium Institute of Astrophysics (MAS); from Proyecto Basal IC\,20009; and from Proyecto FONDECYT Regular \#1141141.
  The authors warmly thank the Kepler Technical and Management Staff for the development, operation, maintenance and success of the Mission.
  \\ {\it Facility:} \facility{Kepler}

\end{document}